\documentclass{aa}
\usepackage{graphicx}
\usepackage{txfonts}
\usepackage[allcolors=blue]{hyperref}

\begin{document}

   \title{Machine learning initialization to accelerate Stokes profile inversions}


   \author{R. Gafeira
          \inst{1,2}
          \and
          D. Orozco Su\'arez\inst{1}
          \and
          I. Mili\'{c}\inst{3,4,5}
          \and
          C. Quintero Noda\inst{6,7}
          \and
          B. Ruiz Cobo\inst{8,9}
          \and
          H. Uitenbroek\inst{5}
          }

   \institute{Instituto de Astrof\'{i}sica de Andaluc\'{i}a (CSIC), Apartado de Correos 3004, E-18080 Granada, Spain\\
              \email{gafeira@mat.uc.pt}
        \and
        Univ Coimbra, IA, CITEUC, OGAUC, Coimbra, Portugal
        \and
            Department of Physics, University of Colorado, Boulder CO 80309, USA
         \and
                    Laboratory for Atmospheric and Space Physics, University of Colorado, Boulder CO 80303, USA
          \and
          National Solar Observatory, Boulder CO 80303, USA
         \and
             Rosseland Centre for Solar Physics, University of Oslo, P.O. Box 1029 Blindern, N-0315 Oslo, Norway
         \and
         Institute of Theoretical Astrophysics, University of Oslo, P.O. Box 1029 Blindern, N-0315 Oslo, Norway
        \and
            Instituto de Astrof\'{i}sica de Canarias, E-38200, La Laguna, Tenerife, Spain.
         \and
         Departamento de Astrof\'{i}sica, Univ. de La Laguna, La Laguna, Tenerife, E-38205, Spain
        }

   \date{accepted 19/02/2021}


  \abstract
    {At present, an exponential growth in scientific data from current and upcoming solar observatories is expected. Most of the data consist of high spatial and temporal resolution cubes of Stokes profiles taken in both local thermodynamic equilibrium (LTE) and non-LTE spectral lines. The analysis of such solar observations requires complex inversion codes. Hence, it is necessary to develop new tools to boost the speed and efficiency of inversions and reduce computation times and costs.}
{In this work we discuss the application of convolutional neural networks (CNNs) as a tool to advantageously initialize Stokes profile inversions. }
   {To demonstrate the usefulness of CNNs, we concentrate in this paper on the inversion of LTE Stokes profiles. We use observations taken with the spectropolarimeter on board the Hinode spacecraft as a test bench mark. First, we carefully analyse the data with the SIR inversion code using a given initial atmospheric model. The code provides a set of atmospheric models that reproduce the observations well. These models are then used to train a CNN. Afterwards, the same data are again inverted with SIR but using  the trained CNN to provide the initial guess atmospheric models for SIR.}
   {The CNNs allow us to significantly reduce the number of inversion cycles when used to compute initial guess model atmospheres (`assisted inversions'), therefore decreasing the computational time for LTE inversions by a factor of two to four. CNNs alone are much faster than assisted inversions, but the latter are more robust and accurate. CNNs also help to automatically cluster pixels with similar physical properties allowing the association with different solar features on the solar surface, which is useful when inverting huge datasets where completely different regimes are present. The advantages and limitations of machine learning techniques for estimating optimum initial atmospheric models for spectral line inversions are discussed. Finally, we describe a python wrapper for the SIR and DeSIRe codes that allows for the easy setup of parallel inversions. The tool implements the assisted inversion method described in this paper. The parallel wrapper can also be used to synthesize Stokes profiles with the RH code.}
   {The assisted inversions can speed up the inversion process, but the efficiency and accuracy of the inversion results depend strongly on the solar scene and the data used for the CNN training. This method (assisted inversions) will not obviate the need for analysing individual events with the utmost care but will provide solar scientists with a much better opportunity
to sample large amounts of inverted data, which will undoubtedly broaden the physical discovery space. }


   \maketitle
%

\section{Introduction}
Inversion techniques have become the most powerful tools for extracting information from observed Stokes profiles regarding the physical conditions of the Sun’s atmosphere. Starting from spectropolarimetric observations of the Sun, the inversion process is based on a non-linear least-squares fit to all four Stokes profiles, which delivers estimates of relevant height-dependent physical parameters such as temperature, velocity, and the vector magnetic field \citep[for a comprehensive review, see][]{LRSP}. In the context of solar physics, these methods are usually applied to high spatial and spectral resolution full-polarimetric observations. Spectropolarimetric inversions were first employed by assuming simple model atmospheres, such as the Unno-Rachkovsky  solution to the transfer of polarized radiation (the radiative transfer equation) in a Milne-Eddington model atmosphere \citep{1982SoPh...78..355L,1987ApJ...322..473S,HAO1987,1990ApJ...348..747L}. Later, \citet{Cobo1992} developed the SIR\footnote{Stokes Inversion based on Response functions} inversion code, which is capable of working with fully stratified atmospheres in local thermodynamic equilibrium (LTE) conditions  (see also SPINOR\footnote{Stokes-Profiles-INversion-O-Routines \citep{2000A&A...358.1109F}}). The SIR code uses analytical response functions that significantly reduce computational times and speed up the inversion process. More recent codes, such as NICOLE\footnote{Non-LTE Inversion COde using the Lorien Engine \citep{NICOLE}}, STiC\footnote{STockholm inversion Code \citep{2016ApJ...830L..30D}}, SNAPI\footnote{Spectropolarimetric non-LTE Analytically Powered Inversion \citep{SNAPI}}, and DeSIRe\footnote{Departure Coefficient Stokes Inversion based on response functions (in preparation)}, brought improvements by, for instance, allowing non-LTE effects and introducing partial frequency redistribution and sophisticated regularization techniques. Introducing new physics into the inversion problem significantly increases the level of realism, but, on the other hand, these new codes are very demanding from a computational resource point of view. The reason is twofold: First, the forward problem itself (i.e. the calculation of the emergent spectrum from a model atmosphere) is far more complicated than in LTE conditions; second, analytical response functions are not always available \citep[but see][]{SNAPI}.

In the advent of new solar telescopes, such as the Daniel K.\ Inouye Solar Telescope \citep[DKIST;][]{DKIST} or the European Solar Telescope \citep[EST;][]{2013MmSAI..84..379C}, the numerical complexity of the inversion process presents an even bigger hindrance. For example, state-of-the-art Fabry-P\'erot based instruments can deliver millions of polarized spectra in a matter of seconds (i.e. several megapixel-sized images taken at different wavelengths and polarization states). On the other hand, the inversion of one spectrum takes from a few seconds up to a few hours of CPU time, depending on the physics involved in the spectral line formation processes. One can easily imagine that for one day of observations with one of these instruments we might need years of CPU time, something that we cannot afford since it is expensive both in terms of funds and time. It is in our best interest to develop faster inversion methods and seek new inversion strategies.

Recently, \cite{paperI} outlined an inversion scheme that completely relies on convolutional neural networks \citep[CNNs; see e.g.][]{CNNref}. Neural networks are powerful machine learning tools that allow us to estimate output parameters from a set of input data that are related in a very complex and non-linear fashion in an incredibly 
short period of time (four orders of magnitude faster). Studies such as \citet{Socas_Navarro_2005},\citet{Socas_Navarro_2005_1}, \citet{radyninv}, \citet{CNN_INV_AA}, and \citet{2019ApJ...875L..18S} have shown the advantages of neural networks over the standard methods when applied to solar data. \citet{CNN_INV_AA} specifically used synthetic spectra calculated from state-of-the-art atmospheric models produced by the MURaM MHD code \citep{MURAM} to train a neural network and then applied the trained network to real data taken with the Hinode spectropolarimeter  \citep{2007SoPh..243....3K,2008SoPh..249..167T,2013SoPh..283..579L,2013SoPh..283..601L}. This neural network was already used, for example, by \cite{2020A&A...637A...1K}, who applied it to Swedish Solar Telescope (SST) observations in a very similar way to the assisted inversion presented in this paper. 

In the case of \citet{paperI}, they first inverted a representative set of synthetic data and then used a CNN to map the relationship between the input (polarized spectra) and the output (values of inferred physical parameters at pre-chosen depth points, hereafter called nodes). They also assumed that there is no spatial coupling between pixels, as in \citet{CNN_INV_AA}. All these methods can be seamlessly applied to either LTE or non-LTE spectral lines provided the training data are available, although we will concentrate here on the LTE case.

In this work we use CNNs to estimate the atmospheric parameters in a similar way to previous studies and following the CNN architecture presented in \citet{paperI}, though with a rather different goal. Undoubtedly, CNNs constitute a very powerful tool for Stokes profile inversions; however, there exist limitations resulting from the lack of physical information in the training set and from the lack of physical connections between the parameters. Hence, CNNs are still in a somewhat early stage of development regarding applications to solar physics, and there might be reason to not fully trust them. Instead, we aim at improving the performance of Stokes inversions on large datasets. Specifically, we use CNNs to estimate optimal initial model atmospheres for each pixel in order to reduce the number of iterations needed by the inversion code to reproduce the observed Stokes profiles. As it turns out, this strategy significantly reduces the computational time. Initial model atmospheres are a key ingredient of the inversions because they are non-convex optimization methods and choosing appropriate initial values of the parameters greatly improves our chances of finding global minima. That is, the fitting process is faster and more robust.

In this paper we discuss the limitations of our CNN model (Sect. 2) in estimating the solar atmosphere from a set of Stokes parameters and introduce a method (Sects. 3 and 4) based on CNNs to improve and accelerate the whole inversion process, with a particular focus on parallel inversions. This work can be seen as a proof of concept of the method, and the results are extendable to other spectral lines, instruments, or scenarios where computational times are the bottleneck of data analysis. The presented method  allows solar scientists to initialize inversions using a pre-computed CNN model (which will be publicly distributed along with this paper), hence reducing the total computational time. We will refer to this methodology as `CNN assisted inversions'. The method has been implemented as an option on a Python wrapper (Sect. 5) that automatically sets up the parallel inversion, simplifying the whole  process for users with little knowledge of CNNs. The wrapper will also allow a parallel inversion to be set up from a variety of instruments, lines, and configurations with or without CNN-based initialization. Conclusions can be found in Sect. 6.


\section{Convolutional neural network}
The goal of this specific CNN is to estimate solar atmospheric parameters from a set of input Stokes profiles (I,Q,U,V). 
In the current approach we neglect any kind of correlation between adjacent pixels coming from, for instance, telescope diffraction or stray light within the instrument. The CNN is therefore based on a 1D convolutional architecture \citep{CNNref} following the strategy presented in \cite{paperI}, which, in turn, is based on the approach in \cite{Parks2018}. In our approach, the inputs are the Stokes spectra, which are treated as 1D arrays (with wavelength dependence) with four layers (for four Stokes parameters). Generally speaking, the output should be a depth-stratified atmospheric model. By atmospheric model we mean: tabulated values of the relevant physical parameters (temperature, velocity, and magnetic field) at each depth point on the given $\log\tau$ grid. However, the output in this case can easily be more dimensional than the input. For most inference techniques, this opens up the possibility of severe degeneracies in the output. Another way to look  at it is that the intrinsic dimensionality of the spectropolarimetric data (i.e. the number of vectors needed to represent the data) is significantly smaller than its formal dimensionality \citep[the number of wavelength points; see e.g.][]{ID}. Standard inversion codes usually simplify the output by using the so-called nodes \citep[e.g.][]{Cobo1992}. That is, even though the fully stratified atmosphere is the output, the number of degrees of freedom is significantly lower as only pre-chosen depth points are allowed to vary freely and the other depth points are found by interpolation. 

For CNN inversions, \cite{paperI} related input spectra to the output node values. \cite{CNN_INV_AA} related input images at an array of wavelengths to the output atmospheres sampled at coarse depth points. In this work, to reduce the dimensionality of the output, we sampled each parameter at ten depth points, which play the role of nodes. After the inference, we used the information from these points to reconstruct, using linear interpolation, the entire atmosphere to the desired optical depth resolution. It is definitely possible to construct a neural network that outputs the fully stratified atmospheres since the up-sampling or interpolation can be made to be a part of the neural network. One example is the work of \citet{radyninv}, who output fully stratified atmospheric models on a very fine grid. Notably, their neural network architecture is much more complicated than ours (an invertible neural network that uses a latent space). We are currently limiting our approach to much simpler architectures.

The global structure of the CNN model is the following: At the beginning, there are two convolutional layers with kernel sizes of seven and five, respectively, each of which is followed by a max-pooling layer with a pool size of two. Then there is a convolutional layer with a size of three followed by a dropout layer with a drop probability of 0.15. The last layers consist of two fully connected layers, one with a dimension of four times the number of spectral points and the other with the total number of atmospheric points, that is, the number of nodes that we seek (typically, six physical parameters for each of the ten depth points). The convolutional layers and the first dense layer use the Rectified Linear Unit (ReLU) activation function \citep{Nair:2010:RLU:3104322.3104425}, and the last dense layer uses a linear function. The model is trained using the optimizer AdaMax \citep{2014arXiv1412.6980K} and a batch size of 64. All the hyperparameters were obtained by trial and error until we found a neural network that offers satisfactory performance. One of the criteria for the satisfactory performance was the ability to use the same architecture for different instruments with different wavelength samplings. Table 1 shows a summary of the used CNN. It should be noted that CNN architecture design in general is not a very rigorous process and there can be significantly different architectures that exhibit similar performances.

Thus, the resulting CNN allows the estimate of the temperature, velocity, and vector magnetic field from a variety of instruments and spectral lines.  This introduces a limitation on the number of layers and kernel sizes that can be used in order to accommodate the few data points typically obtained from filtergraph-based instruments. This particular CNN can be specifically trained and used for several instruments and different sets of spectral lines. More details on the implementation are given in Sect. \ref{warpper}. The general structure of the neural network is very similar to the one used in \citet{paperI}. The main differences are that input now contains all four Stokes parameters and that the output nodes are basically coarse atmospheric stratifications instead of the nodes used by the inversion code.

Within the CNN, most quantities are internally normalized using given scaling factors (see Table ~\ref{table1}). Moreover, the input Stokes parameters are normalized to the local quiet Sun mean intensity. Internally, the CNN gives an extra weight of five to Stokes V (relative to Stokes I) to increase the sensitivity of the parameter.
\begin{table}[h!]
\centering
 \begin{tabular}{||c|c|c||} 
 \hline
  Layer & Size&Activation\\
 \hline
  \hline
  1D convolutional &  7&ReLU\\
 \hline
  1D MaxPooling &  2&\\
 \hline
  1D convolutional & 5 & ReLU\\
 \hline
   1D MaxPooling & 2 &  \\
 \hline
   1D convolutional & 3 & ReLU\\
 \hline
   Dropout & factor of 0.15 &\\
 \hline
   Fully connected  &  4 times the spectral &\\ 
   dense layer &points & ReLU\\
 \hline
    Fully connected & total number of &\\ dense layer& atmospheric points & linear \\
 \hline
\end{tabular}
\caption{Proposed CNN model architecture.}
\label{cnn}
\end{table}

\begin{table}[h!]
\centering
 \begin{tabular}{||c | c c c c c ||} 
 \hline
  & Temp. & Velocity & B & $\gamma$ & $\chi$ \\ [0.5ex] 
 \hline
 Units & Kelvin & cm/s & Gauss & degree & degree \\ 
 \hline
 Norma. & 20\,000 & 300\,000 & 6000 & 180 & 360 \\[1ex] 
 \hline
\end{tabular}
\caption{Normalization factors applied to different physical quantities within the CNN.}
\label{table1}
\end{table}

\section{Inversion procedure and CNN training data}
\label{dtip}
\begin{figure}
\centering
\includegraphics[width=\hsize,trim={1.5cm 2.5cm 0cm 2.5cm},clip]{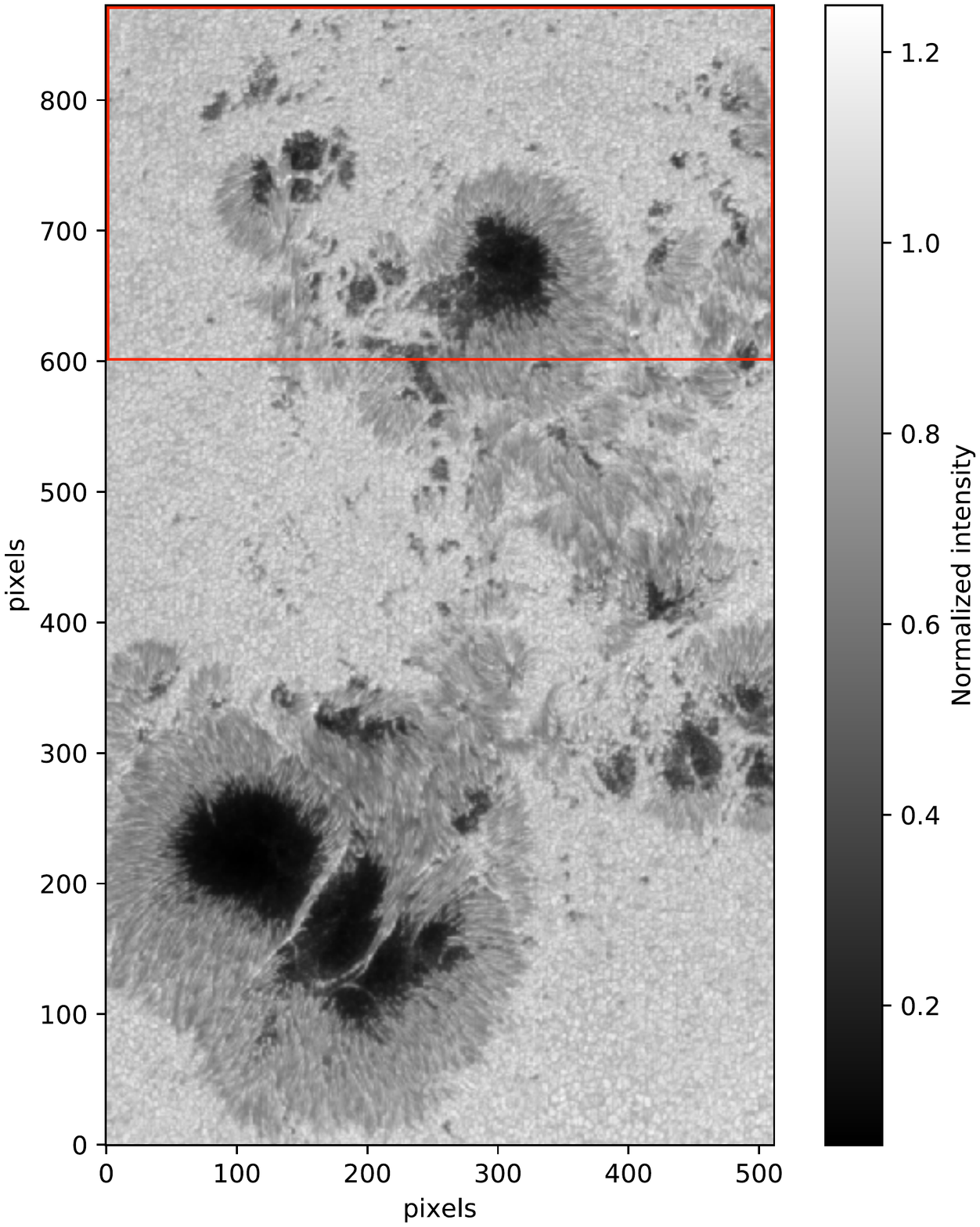}
  \caption{Continuum intensity image of NOAA 12192, normalized by the quiet Sun mean intensity,   taken by the Hinode spectropolarimeter on 25 September 2014. Pixel size is 0,16. The red box represents the field-of-view considered for training the CNN in case 1.}
     \label{plot_full}
\end{figure}

The employed data were taken by the spectropolarimeter on board the Hinode spacecraft in the Fe I 6301.5 and 6302.4 \AA{} lines. They cover part of Active Region NOAA 12192 (Fig.~ \ref{plot_full}). We selected these data because they show a variety of solar structures (i.e. several pores, fully developed sunspots, and quiet Sun regions), and therefore different physical scenarios coexist within the same dataset.  

As explained before, the goal of this study is to use a CNN to generate initial (guess) model atmospheres for the Stokes inversion. To evaluate the performance of such an assisted inversion and of the CNN itself, we performed three different inversions using the same data. 

The first inversion, which we will refer to as
the `reference set', was used to analyse the performance of the assisted inversion. The second inversion was our `training set', where we used a few node positions from the reference set taken at different heights along the atmospheric parameters to train the CNN and to estimate its performance (see Sect.~\ref{nnd} for more details).
The third inversion, the `second reference set', was used to estimate the intrinsic variations (errors) on the final inversion results caused by using different initial guess models. In all three cases, we used the SIR code, which works in LTE \citep{Cobo1992}.

\begin{figure*}
\centering
\includegraphics[width=\textwidth]{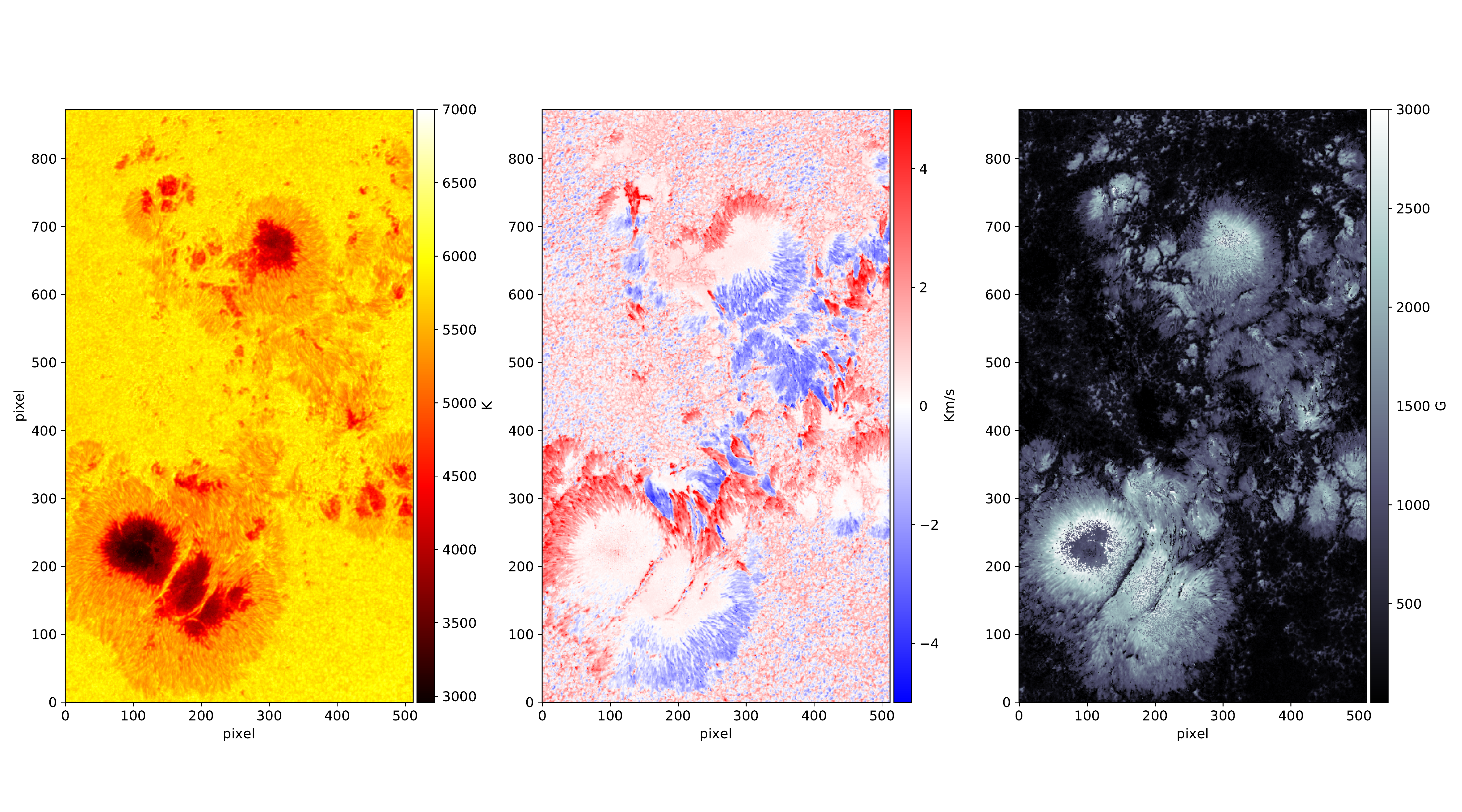}
\includegraphics[width=0.67\textwidth]{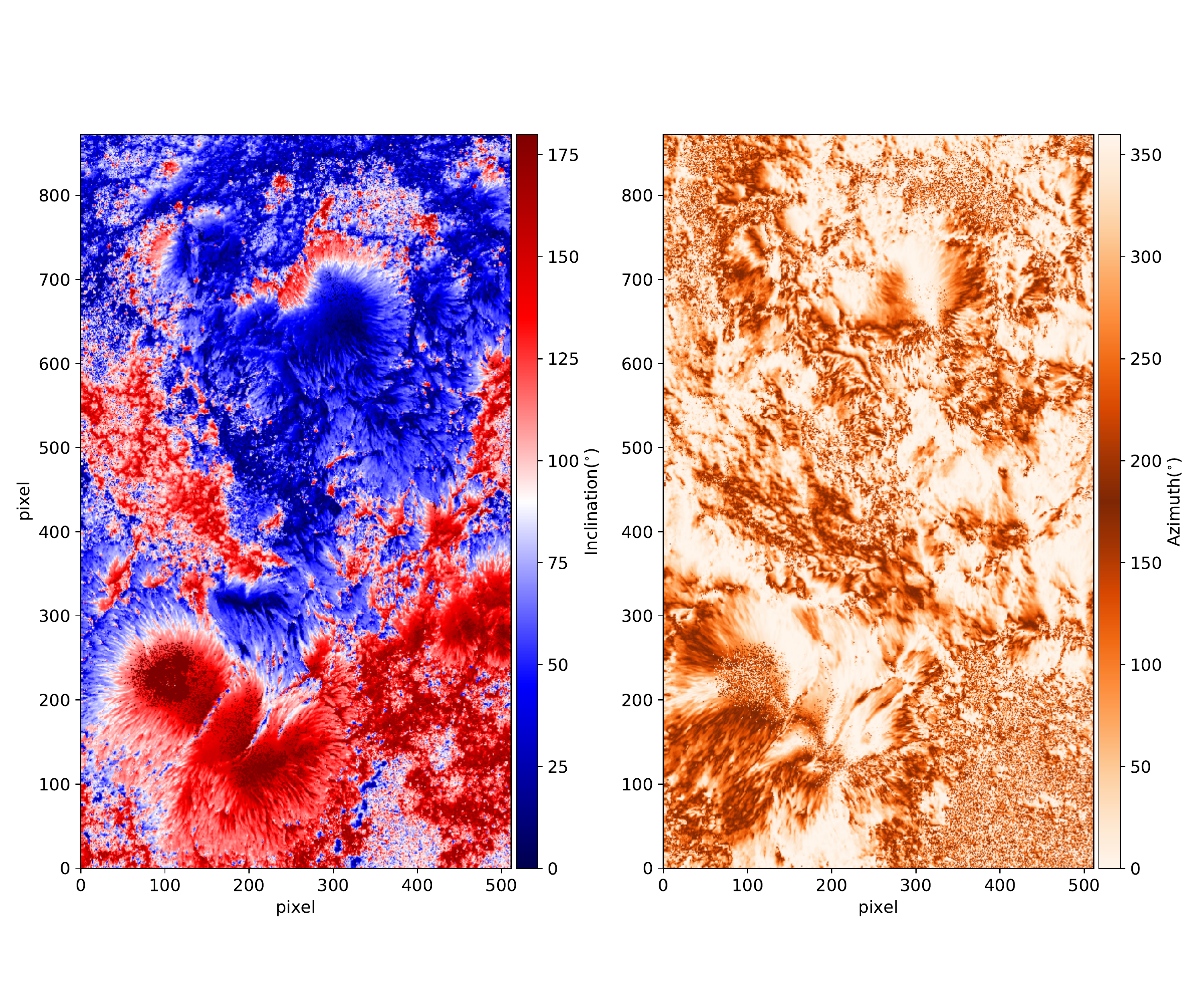}

\caption{Temperature, velocity, magnetic field, inclination, and azimuth maps inverted using a classical inversion approach. The reference set for the inversion strategy applied in this case is given in Table \ref{table3}.}
\label{in_inv}
\end{figure*}

For the reference set, we used five nodes for the temperature, three nodes for the magnetic field strength and velocity, and
 one node for the field inclination, field azimuth, and microturbulence. The macroturbulence velocity was also inverted. Thus, we have a total of 15 free parameters. The number of wavelength points across the line is 112 per Stokes parameter. Figure~\ref{in_inv} displays the results for the temperature, the line-of-sight velocity and the magnetic field strength, inclination, and azimuth at optical depth unity. This model fits the Stokes profiles to a level we are comfortable with, showing very low $\chi^2$ values (with a mean value of 0.012); that is, the rms errors are close to the noise level, on average. It is important to stress that the inversion itself is not performed at once. Rather, the inversion code starts with a lower number of nodes and then increases them in a series of `cycles' as if we were inverting the same pixel several times while increasing the complexity of the model (i.e. smoothly increasing the number of free parameters). This allows the inversion code to reach the best solution at the expense of computational time. In our case, the number of cycles is set to three. The full description of the nodes per cycle is given in  Table \ref{table3}.

\begin{table}[h!]
\centering
\small
 \begin{tabular}{|c|cccc|} 
 \hline
& reference & training  & assisted  & 2\textsuperscript{nd} reference\\
& set  &  set &  inversion &  set\\

 \hline
 Nº of cycles & 3 & 3 & 1 & 3 \\ 
 \hline
  &  &  &  &  \\ 
 \hline
 Nº of nodes &  &  &  &  \\ 
 \hline
 Temperature & 1,3,5 & 1,3,5 & 5 & 1,2,4 \\ 
 \hline
  Velocity & 1,3 & 1 & 3 & 1,2 \\ 
 \hline
  Field strength & 1,3 & 1 & 3 & 1,2 \\ 
 \hline
  Field inclination & 1 & 1 & 1 & 1 \\ 
 \hline
  Field azimuth & 1 & 1 & 1 & 1 \\ 
 \hline
   Microturb. & 1 & 1 & 1 & 1 \\ 
 \hline
    Macroturb. & 1 & 1 & 1 & 1 \\ 
 \hline
\end{tabular}
\caption{Number of cycles and number of nodes per cycle for each different inversion.}
\label{table3}
\end{table}

For the training set the setup consisted of five nodes in temperature and one node in velocity, microturbulence, and magnetic field strength, inclination, and azimuth . Macroturbulence was also inverted. We again used three cycles (Table \ref{table3}). At this point, one could wonder why we used a simpler model to train the CNN and not a more sophisticated one that successfully reproduces the observed Stokes profiles. The reason is twofold: first, to reduce the degeneracy of the fit caused by the high number of nodes. This directly translates into a decrease in the accuracy of the CNN performance. Second, SIR inversions are much more robust and also stable when inverting the Stokes parameters with no height variations in the line-of-sight velocity or the vector magnetic field. The inferred atmospheric models would not be able to reproduce the asymmetries exhibited by the Stokes profiles in this scenario. However, we recall here that the goal is to generate initial guess models for the inversion rather than attempt to perfectly reproduce the observations using the CNN.

For the second reference set we used a maximum of four nodes for the temperature, two nodes for the magnetic field strength and velocity, and one node for microturbulence, the magnetic field inclination and azimuth. The macroturbulence velocity was also inverted. Apart from using a different number of nodes along the atmosphere, this inversion differs from the reference set in the fact that a different initial model atmosphere was used. For the second reference set, we used the FALC model atmosphere as the initial guess, while for this set we used the HRSA model atmosphere \citep{1971SoPh...18..347G}. (In the second reference set, the mean $\chi^2$ value is 0.02). The results then allowed us to analyse uncertainties ascribed to the inversion code itself. 

 It is important to emphasize that the numbers of nodes and cycles are optimized in such a way that the Stokes inversion of this particular pair of lines provides good results for most pixels in the data, on average. Should the user need a more complex atmosphere to reproduce particular physical imprints in the Stokes parameters, the numbers of nodes and cycles have to be changed accordingly to accommodate the extra complexity. As a matter of fact, for the particular case of non-LTE lines, the number of nodes necessary to have a reasonable training set or even convergence  may need to be increased due to the more complex atmospheric structure. The CNN model presented in this paper can be easily tailored to accommodate such cases.

\section{CNN training strategy and assisted inversion results}
\subsection{CNN training}
As previously explained, we inverted the data following two different approaches whose main difference lies in the number of nodes taken along the different physical model parameters: the training set and the reference set. Both inversions were carried out for the full map. Here we analyse the influence of the choice of the training data on the CNN performance, that is, on whether we used the whole dataset or only parts of it to train the CNN. Hence, we trained the CNN using three different approaches.
The only thing in common between them is that we used 200 epochs\footnote{An epoch is defined as a full cycle of training of the forward estimation and back propagation.} for training the CNN.

In particular, we considered two scenarios. For case 1, we trained the CNN just using part of the data, in particular the upper part of the map (red box in Fig. \ref{plot_full}, from row 600 on).
For case 2, we trained the CNN using the full field-of-view in Fig.~\ref{plot_full} (i.e. the full training set).

   \begin{figure*}[!http]
   \centering
   \includegraphics[width=0.5\textwidth]{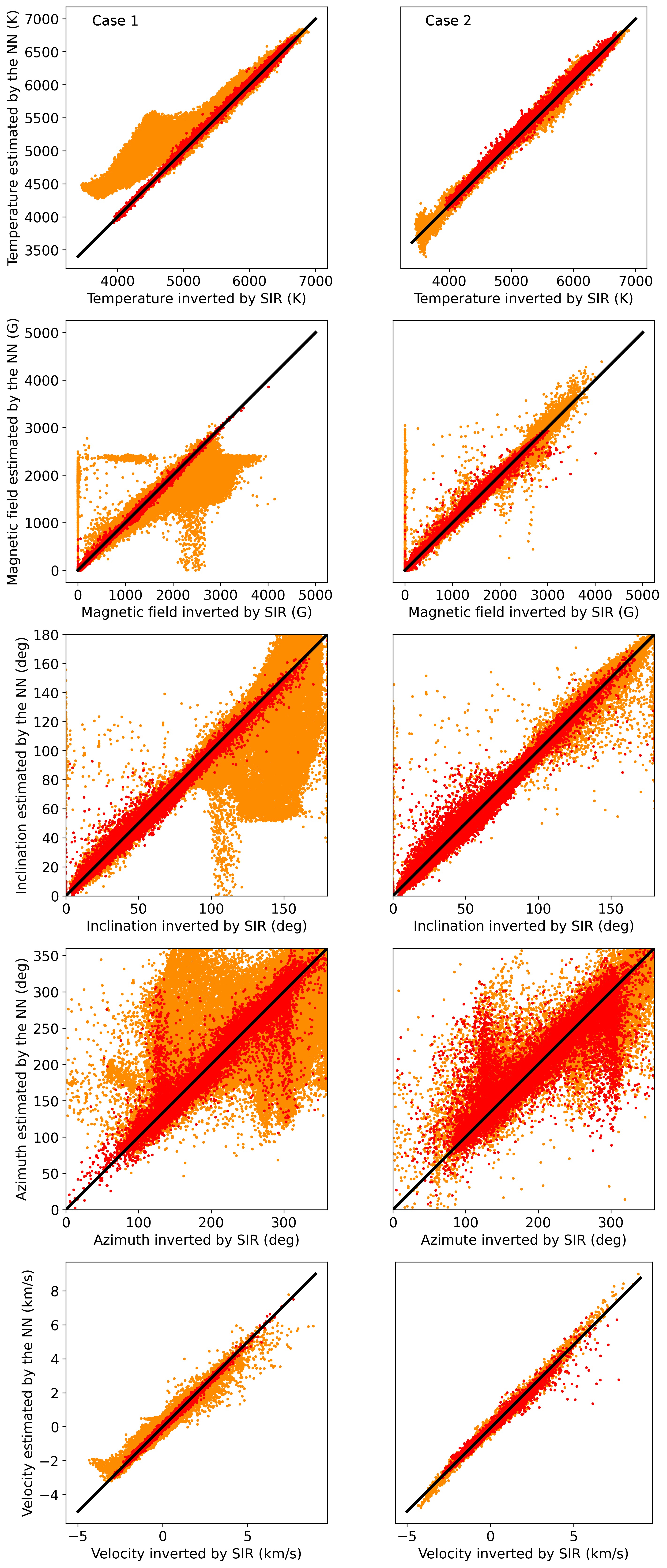}
      \caption{Comparison between the CNN estimation of the temperature, the line-of-sight velocity, and the magnetic field strength, inclination, and azimuth over the whole map using the different training scenarios and the reference inversion. The red dots correspond to pixels present in the red box in Fig. \ref{plot_full}. The orange dots correspond to pixels outside the red box.}
\label{figbig}
   \end{figure*}

The scatter plots in Fig.~\ref{figbig} display a comparison between the inverted atmosphere with SIR (i.e. the training set) versus the CNN estimation for temperature, velocity, magnetic field strength, magnetic field inclination, and magnetic field azimuth at $log(\tau)=0$. In these plots, the influence of the different training scenarios on the results provided by the CNN can be clearly seen. As can be seen in the top panels, which correspond to the temperature stratification, the performance of the CNN estimate over the full field-of-view is worse in case 2 (right panel), where only part of the map was used to train the CNN, than that for case 1 (left panels). Even with the presence of several solar structures within the red box, the information contained in the top part of the map is not sufficient to properly train the CNN and then use it to infer the physical parameter for the rest of the map.

A closer look at the scatter plots reveals that, for the temperature case, the CNN is able to properly retrieve the temperature values in the region that we used to train the network, represented in red. For regions outside the training area, represented in orange, the CNN is not performing as required. The deviations are mainly concentrated in the low-temperature region. For higher temperatures, mostly granulation, the temperature distribution is closer to the initial input. This means that the lack of information was important enough in the training process, but it also demonstrates how the CNN is able to be updated and learn from new datasets, improving its performance. The top-right panel shows the results corresponding to case 2 for temperature. In this case the CNN is able to retrieve, with good accuracy, the initial training values, with rms values below 40 Kelvin over the whole range of temperatures. 

Regarding the vector magnetic field, the behaviour of the CNN with respect to the training set is similar to that for the temperature. Case 1 fails to reproduce the stronger fields and their inclinations and azimuths. In this particular case the reason is that the training data do not cover the large variety of magnetic field configurations present in the full map. Indeed, the sunspots of the lower part show opposite polarity. It is also worth noticing that there are points in the magnetic field strength scatter plot that show field strengths close to zero as a result of the SIR inversion, while they are not zero in the CNN estimate. After carefully looking at this issue, we found that this is caused by a bad convergence of SIR in the darkest parts, which correspond to the pores and sunspot umbral areas. Due to the low temperature in the centre of the umbra, the contribution of molecular transitions to the observed spectra becomes significant, limiting the accuracy of the fits. This effect, which is also noticeable in the magnetic field inclination and azimuth maps, could be mitigated by including molecular lines in the inversion process. In the case of the CNN inversion results, we can see a significant number of umbral and pore pixels where the magnetic field is now much larger than zero. This indicates that the CNN may be less sensitive to the contamination of molecular lines. In this particular case of assisted inversion, we can have better, or more coherent, results than the reference set. In our interpretation, this is because the CNN initialization can keep the inversion code from falling in into a local minimum, caused by the molecular lines, and then not being able to move to another solution.

Similar results can be found when comparing the line-of-sight velocity inferred with the CNN and the training set.\ In other words, the CNN has larger uncertainties when trying to estimate the velocities of the map not `seen' during the training process.

Overall, the model obtained in case 2, using the full map to train the CNN, shows a better performance in estimating the physical parameters. For this reason, we used it to compute the initial model atmosphere that is input into the SIR inversions.

\subsection{Assisted inversion strategy}

In this section we present the assisted inversions and the tests run to estimate their performance. These tests focused on the differences between the results of the standard inversions and the assisted inversions using a CNN model trained on the same dataset. The results of the assisted inversion when a CNN model trained on completely unseen data is used to initialize the inversion were also analysed.

\begin{figure*}[h]
   \centering
   \includegraphics[width=\textwidth]{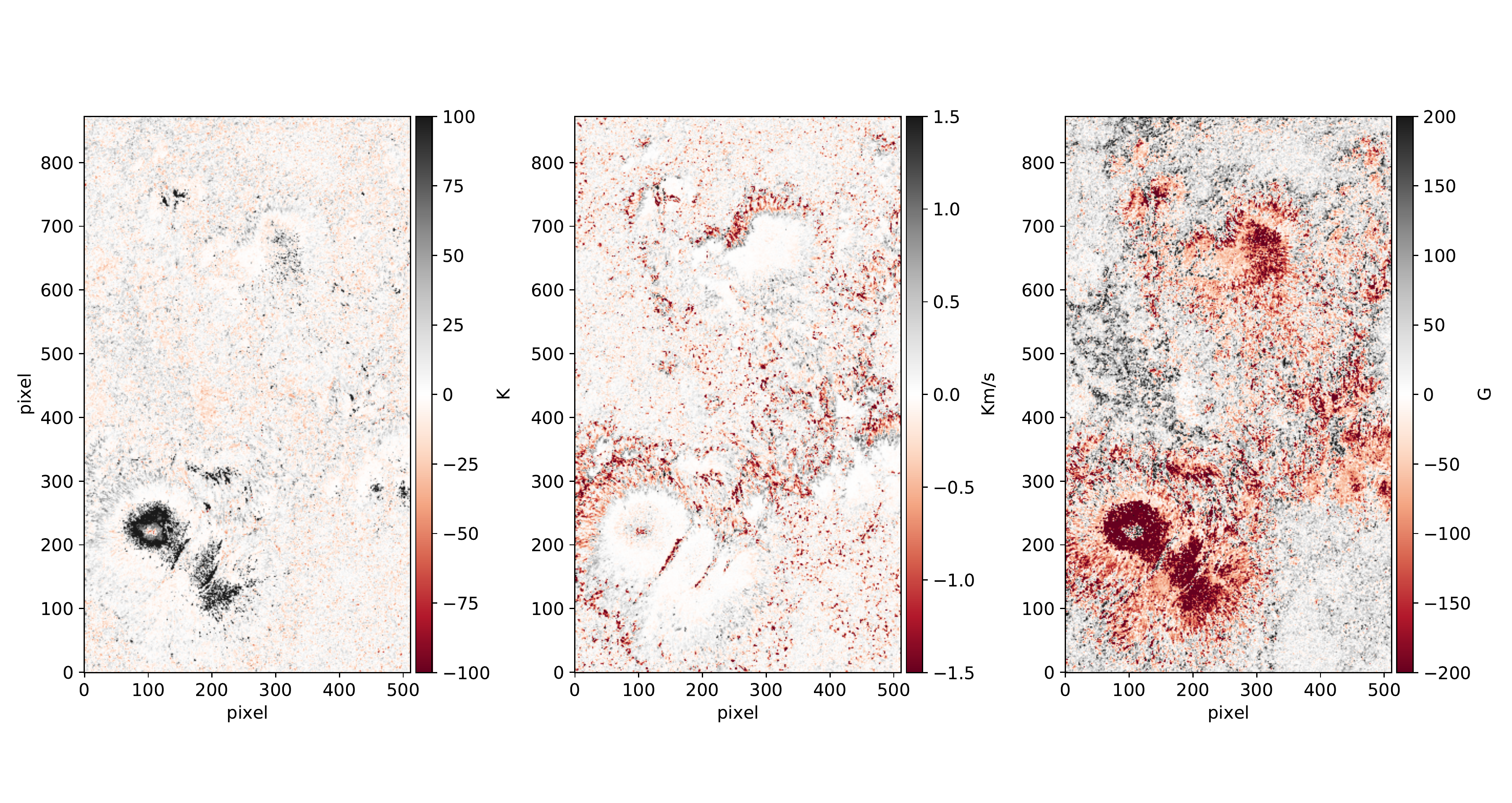}
   \includegraphics[width=0.7\textwidth]{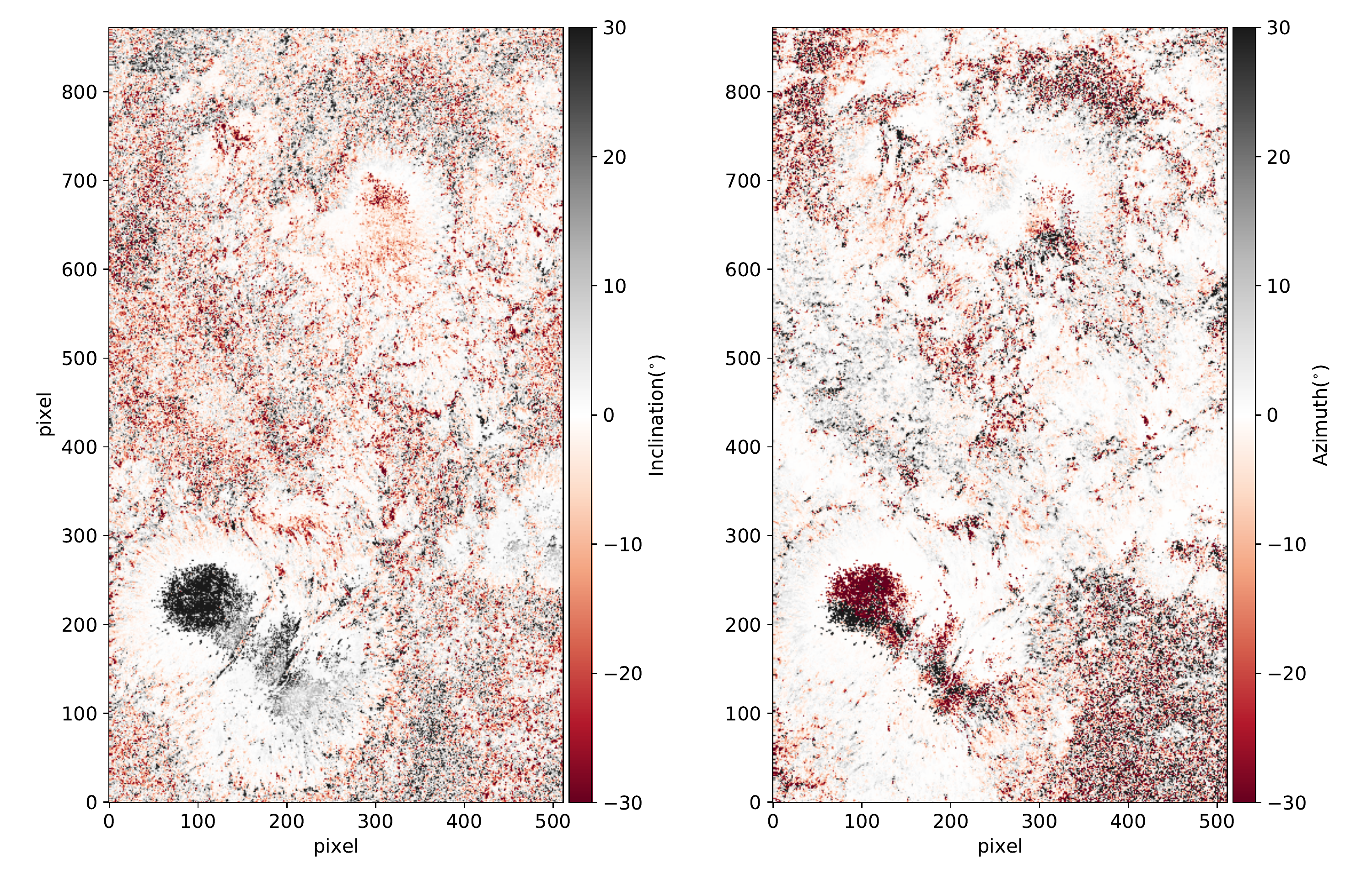}
      \caption{Difference between the reference set and the assisted inversion results for the temperature, the velocity, and the magnetic field strength, inclination, and azimuth.}
         \label{as_inv}
   \end{figure*}

The assisted inversion consists of a standard inversion in which we used the CNN to estimate optimum initial guess model atmospheres. With this approach we expected to start the inversion from an atmospheric model that is fairly close to the final one. This strategy allowed us to reduce the number of cycles needed to run the full Stokes inversion. To test it, we inverted the map using just one cycle and the same number of nodes used in the last cycle of the reference set. Using just one cycle, and not three as in the reference set inversion, reduced the computation time in this particular case by a factor of approximately two.
In Fig.~\ref{as_inv} we show the differences between the reference set and the assisted inversion results for the different atmospheric parameters at $\log \tau =0.0$.

\begin{figure*}[h]
\centering
\includegraphics[width=0.33\textwidth]{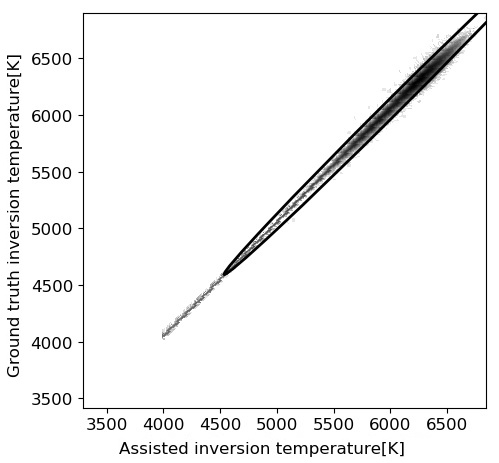}
\includegraphics[width=0.323\textwidth]{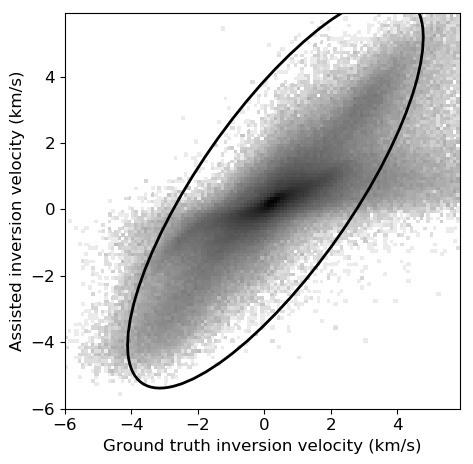}
\includegraphics[width=0.33\textwidth]{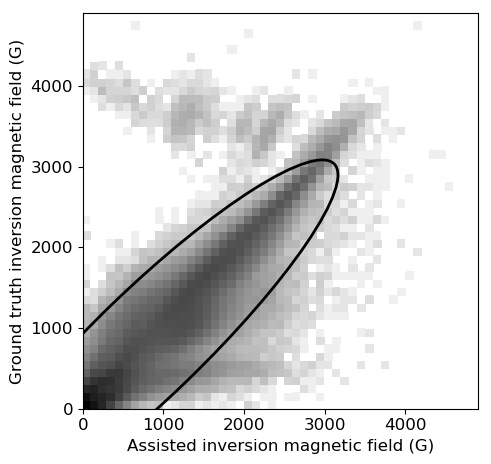}
\caption{Log scale of the 2D histogram of the temperatures, magnetic field strengths, and velocities obtained by the reference set inversion and the assisted inversion. The curved black line indicates the 99\% confidence level.}
 \label{t_as}
\end{figure*}

In order to estimate the accuracy and performance of the assisted inversion method, we directly compared the computed the temperature, the velocity, and the magnetic field strength, inclination, and azimuth at $\log \tau =0$ obtained by this inversion with the reference set. Figure \ref{t_as} (left panel) shows the 2D histograms, in logarithmic scale, of the two temperature distributions obtained by the two methods and the 3$\sigma$ confidence ellipse (black curve). The correlation factor is very high, showing that the assisted inversion obtains a solution very close to the reference one. 
\begin{table}[h]
\caption{Pearson correlation factor between the assisted and reference set inversions and between the reference set and the second reference inversion for each atmospheric physical parameter.}
\label{table_corr}
\begin{tabular}{ccc}
 Atmospheric & Correlation & Correlation\\ 
   physical parameter & Assisted inversion & 2\textsuperscript{nd} ref. inversion\\ \hline
 Temperature& 0.99&0.99\\
 Velocity& 0.84&0.85\\
 B strength &0.91& 0.90\\
 B azimuth & 0.79& 0.81\\
  B inclination&0.85& 0.92\\ 
\end{tabular}
\end{table}

Similarly, the velocity distribution is more spread out, as we can see in the 2D histogram in Fig.~\ref{t_as} (right panel) and Table \ref{table_corr}. From a more detailed analysis of the velocity maps obtained by the reference set and assisted inversion maps, we see that most of the pixels that show higher differences are located, once again, in the centre of the umbrae as well as in some penumbra regions. The latter can be explained by the more complex atmospheric structure present in the penumbrae that produces a higher degeneracy level in the fitting process. Ideally, one should use a two-component model to properly fit these regions \citep{2004A&A...427..319B}.

The same applies to the magnetic field intensity, as can be seen from the 2D histogram in Fig.~\ref{t_as} (middle panel). From the analysis of the histogram, we can see a branch of pixels that is separated from the ideal case. This is mainly caused by the pixels in the centre of the umbrae where the inversion code cannot properly fit the Stokes profiles due to the contribution of the molecular lines. The same effect is present in the magnetic field inclination and azimuth (not shown). These last two parameters are typically more difficult to estimate via inversion codes due to the fact that these signals (i.e. Stokes Q and U) are more affected by noise. Additionally, large differences in the azimuth are typically more significant because of the azimuth ambiguity. These effects contribute to the decrease in the correlation between the two distributions.

\begin{figure*}[h]
\centering
\includegraphics[width=\textwidth]{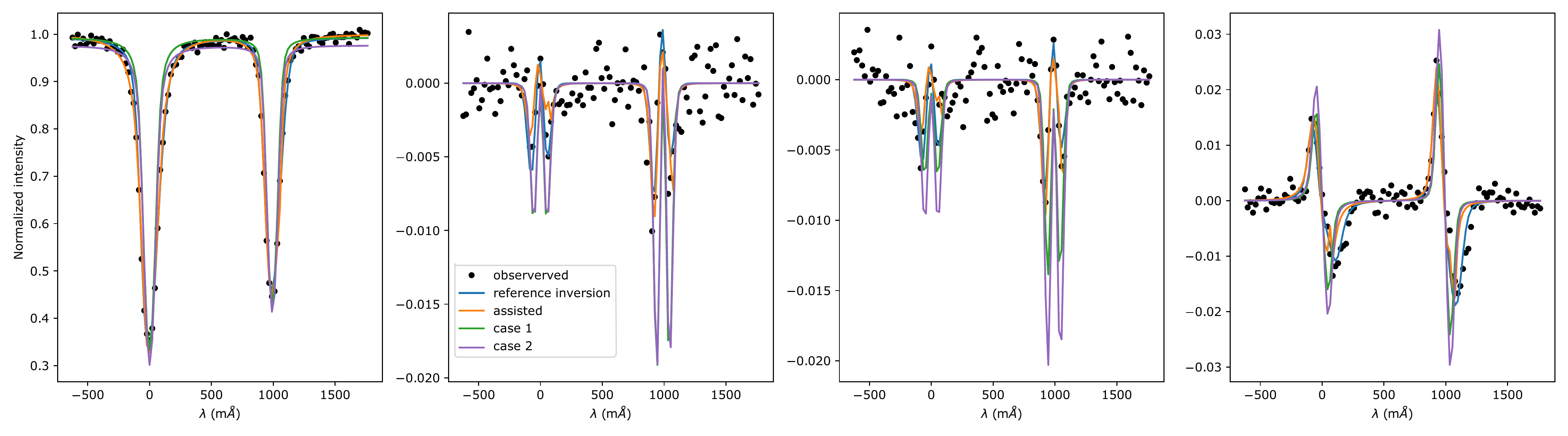}
\includegraphics[width=\textwidth]{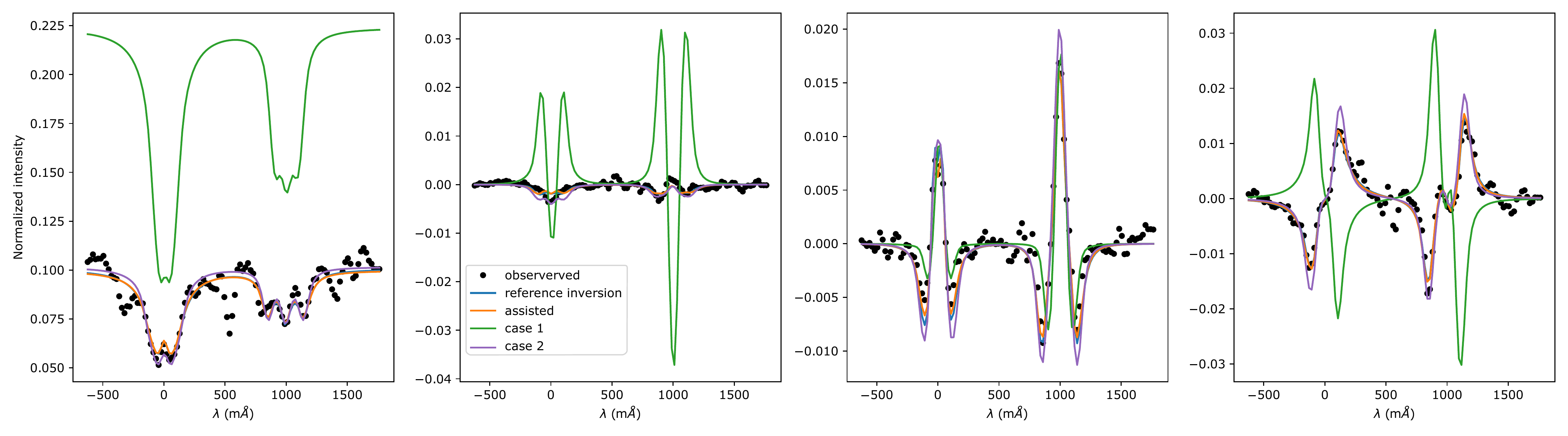}
\caption{Observed and synthesized spectra from the different computed solar atmospheres from a pixel taken from a quiet Sun region and a pixel taken from the umbra.}
 \label{spectra}
\end{figure*}

To better see the differences between the CNN (including the three different training scenarios, i.e. cases, discussed above), the assisted inversion, and the reference set, Fig.~\ref{spectra} shows synthetic profiles computed from the atmospheric models obtained by the different methods. In particular, we concentrate on a pixel taken around a quiet Sun area showing significant Stokes Q, U, and V signals as well as on a pixel taken from one of the sunspot umbrae located in the lower part of the map.

It can be readily noticed that, in almost all the cases, the quiet Sun profiles are fitted successfully. A closer look at the Stokes I fits reveals that both the reference and the assisted inversion fits are almost indistinguishable. The largest differences between these two inversions are in the red wing of Stokes V, although almost at the level of the noise. The fits from the CNN (in the three cases) are also very similar, but fail slightly in reproducing the wings of Stokes I and the amplitude of the linear polarization signals. This behaviour is common in most pixels in the data. The umbral profile example shows much larger differences. Again, the reference and the assisted inversion show almost indistinguishable results. However, there are clear differences between the three different cases run with the CNN. Case 1 (which only included part of the data for training the network) completely fails in reproducing the profiles. Indeed, the Stokes V polarity is completely wrong. Finally, when we used the whole dataset to train the CNN (case 2), the fits were much better. This proves what was discussed before. When the CNN is not properly trained (case 1) it fails to infer the atmospheric models from pixels that are not properly sampled by the training set.

The last step we took was to identify if the differences between the two inversions were induced by the method described in this paper or result from the inversion procedure itself. Therefore, we performed a final inversion with a different initial condition and configuration of the nodes, which gave similar $\chi^2$ values as the reference set. As a reminder, the full description of the nodes per cycle is given in Table \ref{table3}. We then compared the results from the two inversions and computed the correlation factor (see the right column in Table~\ref{table_corr}).  From the analysis of these results, we see that the dispersion for each atmospheric parameter obtained with a second inversion is very similar to those obtained by the assisted inversion. This indicates that the differences between the assisted inversion and the reference set can be explained as variances introduced by the uncertainty in the inversion method itself, which is caused by the nature of line formation, where the contribution from different layers of the atmosphere needs to be taken into account. This causes different combinations of physical parameters along the atmosphere result in the same line profile. As a consequence, this induces dispersion in the derived parameters when compared directly at the same height.

\begin{figure*}
\centering
\includegraphics[width=\hsize]{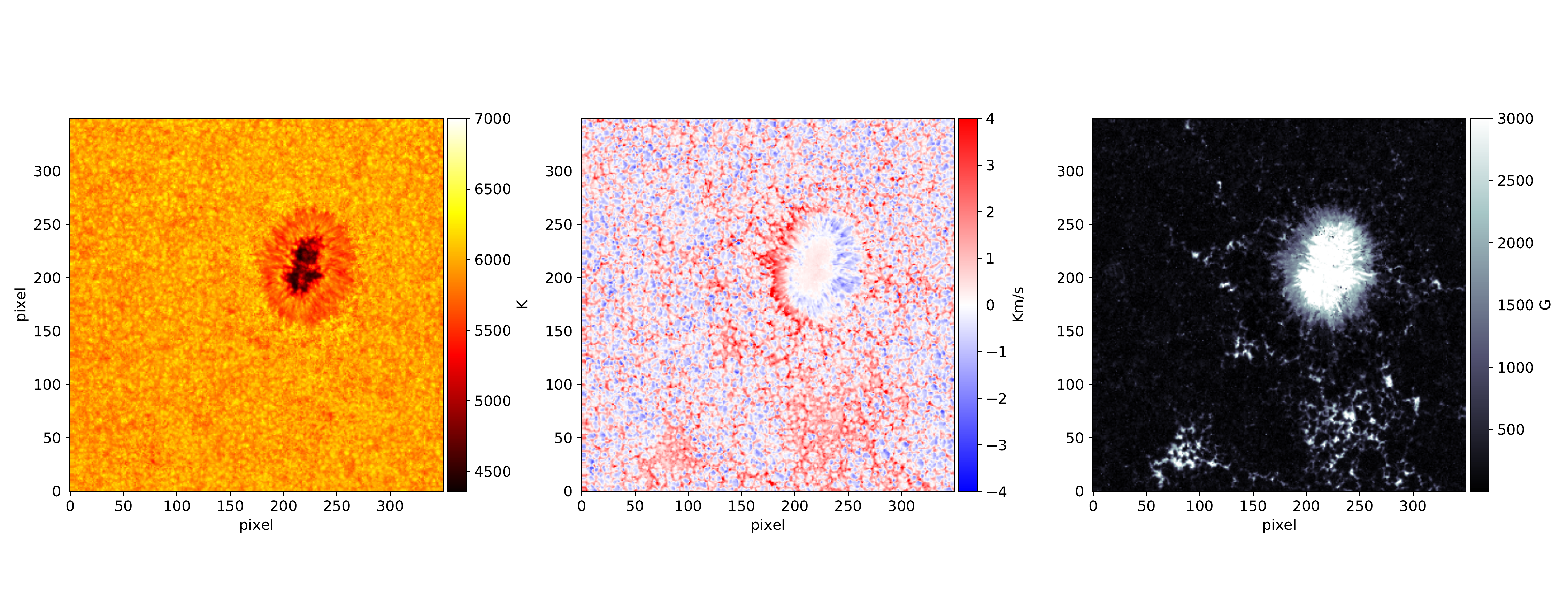}
\includegraphics[width=0.67\hsize]{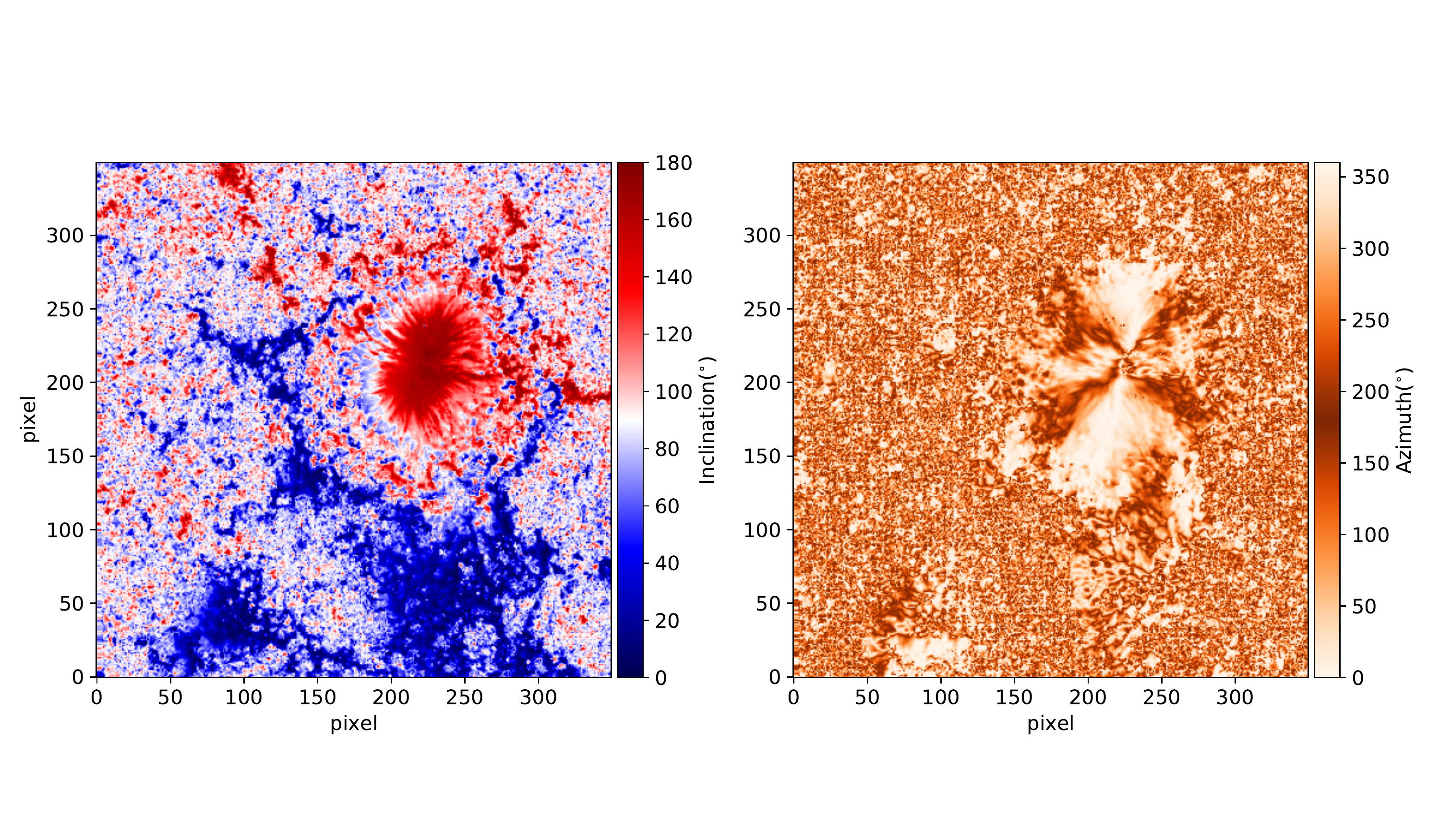}

\caption{Temperature, velocity, and magnetic field strength, inclination, and azimuth maps inferred from the SIR assisted inversion for the CNN unseen active region Hinode map.}
 \label{qs_map}
\end{figure*}

\begin{figure*}[h]
\centering
\includegraphics[width=0.32\textwidth]{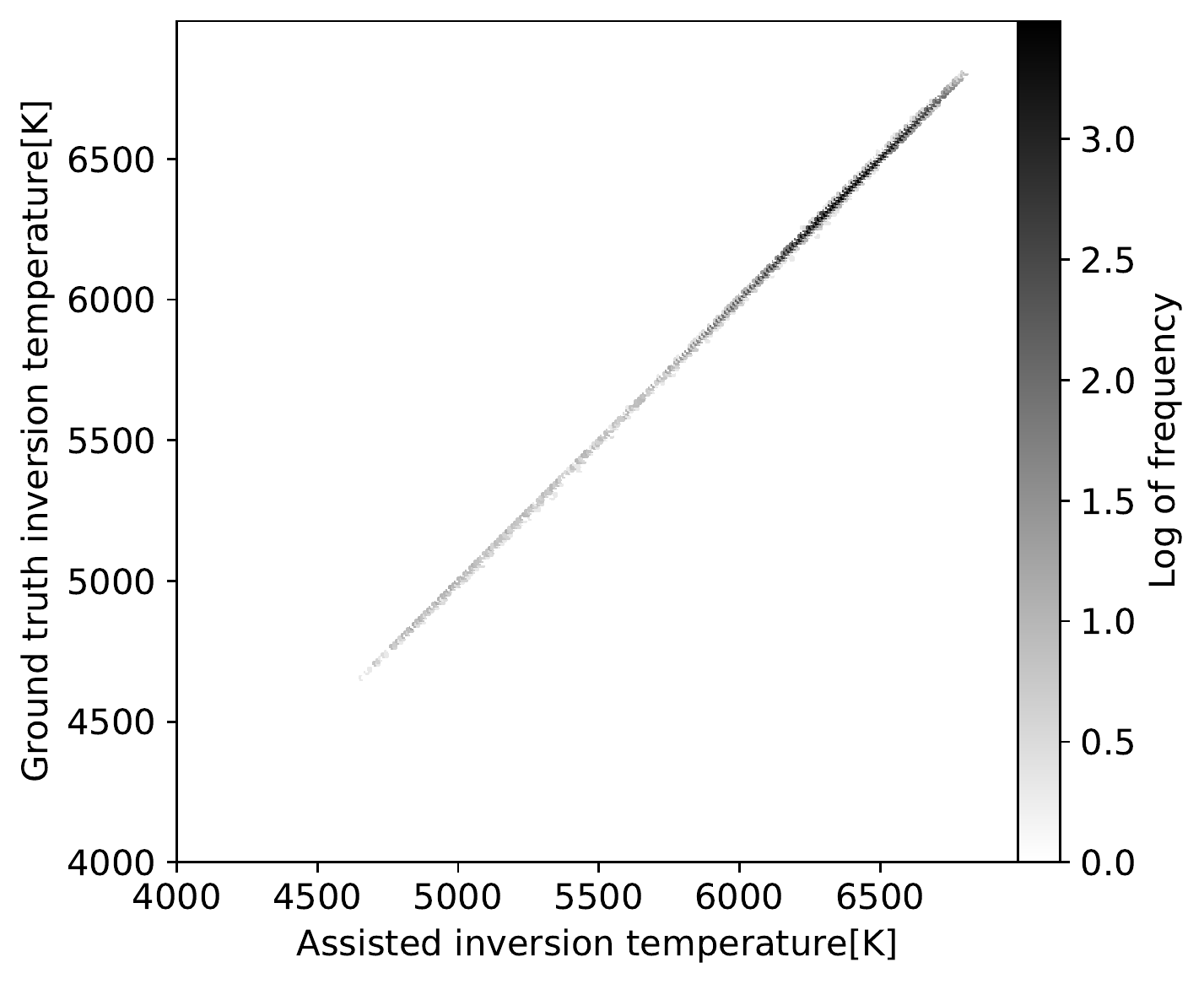}
\includegraphics[width=0.32\textwidth]{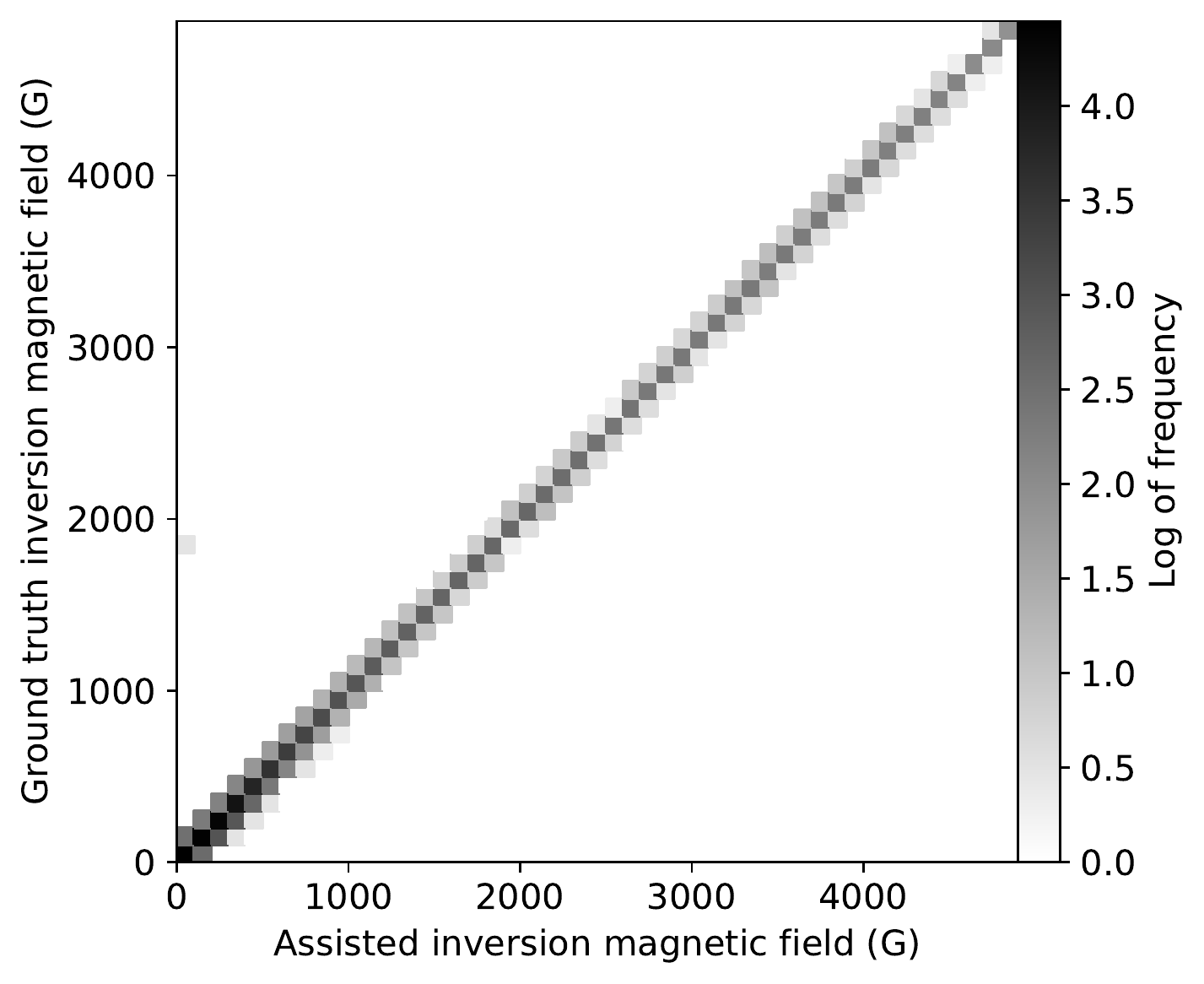}
\includegraphics[width=0.32\textwidth]{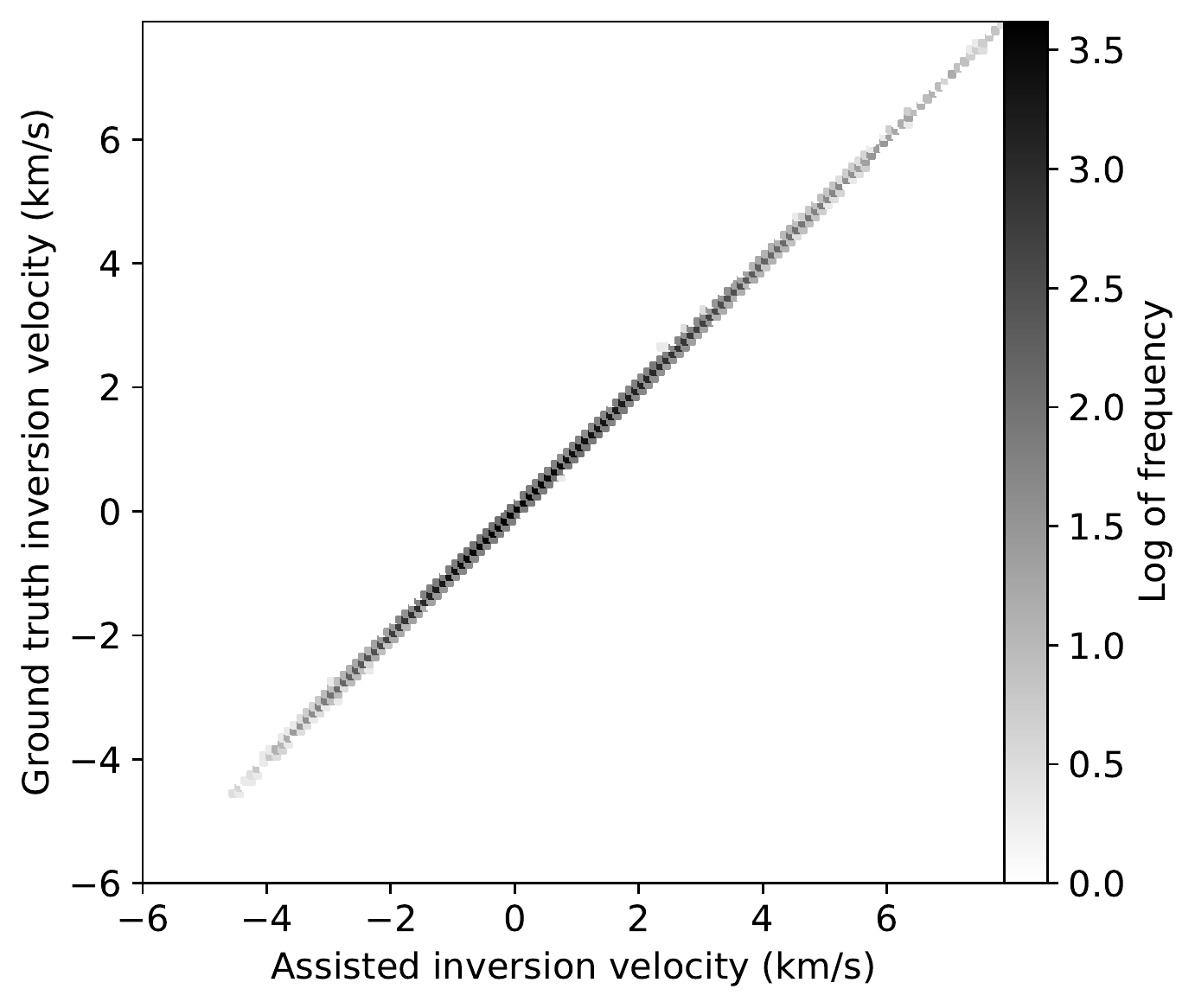}

\caption{Log scale of the 2D histogram of the temperatures, magnetic field strengths, and velocities obtained by the reference set inversion and the assisted inversion.}
 \label{qs_as}
\end{figure*}

\
To evaluate the performance of the assisted inversion when applied to a completely different
and unseen map that was not used for training the CNN, we inverted another Hinode active
region map taken with the same observation configuration on 28 August 2007, which covered
active region NOAA 10969 at $\mu =0.77$. In this case, we performed two
inversions, one to get the reference set (computation time 1.25 hours) and a second assisted inversion (computation time 19 minutes), using only the case 2 CNN model trained using the dataset from Fig. \ref{plot_full} as the model for the assisted inversion. The configuration of both the reference set and the assisted inversion is the same as on the previous test. More details are given in Table \ref{table3}.
This indeed shows us that it is possible to use a pre-computed model derived from a different dataset for assisted inversions. We show the resulting plots (similar to Figs. \ref{in_inv} and \ref{t_as}) in Figs. \ref{qs_map} and \ref{qs_as}.
Table \ref{table_corr_qs} shows the Pearson correlation factor between the reference inversion and assisted inversion result for the Hinode  unseen active region map.

\begin{table}[h]
\caption{Pearson correlation factor between the assisted and reference set inversions for each atmospheric physical parameter.}
\label{table_corr_qs}
\begin{tabular}{cc}
 Atmospheric & Correlation\\ 
   physical parameter & Assisted inversion \\ \hline
 Temperature& 0.999\\
 Velocity& 0.999\\
 B strength &0.999\\
 B azimuth & 0.998\\
  B inclination&0.996\\ 
\end{tabular}
\end{table}

Comparing the assisted inversion results obtained for the dataset where the CNN was trained to the unseen dataset, Figs. \ref{plot_full} and \ref{qs_as} respectively, we see that the latter shows significantly better results, not just on the direct comparison of the atmospheric parameters, but also in the time reduction (with the standard inversion serving as a reference).
The possible explanation for this behaviour can be split into two parts. One is related to the overall simplicity of the profiles. In this dataset, we do not have the very cold umbral regions we have in the first dataset -- we note that those pixels were creating most of the results in the inversion -- and we do not have such intricate penumbral regions. The second possible cause is that in 2009 a Hinode  antenna failed and the data were automatically spatially binned before being sent. While the nominal spatial resolution was the same, the difference in sampling might have influenced the complexity of the spectra and corresponding atmospheres.

In this example we used Hinode data to test the assisted inversions. The strategy works in many other situations, that is, with different spectral lines (either LTE or non-LTE) or instruments such as VSP/DKIST. To prove this, we show (in Appendix \ref{ap2}) the result of assisted inversions applied to observations taken with the GRIS polarimeter  \citep{2012AN....333..872C} installed at the Gregor telescope \citep{2012AN....333..796S} at the Izaña Observatory (Canary Islands, Spain) in the Fe I 1564.7 and 1564.8 nm lines.  The followed strategy was exactly the same, that is, we first trained the CNN with the results from the full SIR Stokes inversion and then re-inverted the data with the assisted method presented here  (i.e. calculating the initial guess model atmosphere using the CNN and reducing the number of cycles in the inversion). The gain in speed is significant, taking less than half of the time to invert the full map (excluding the CNN training time, which took approximately 5 hours). Should the user want to invert new Gregor datasets (taken in a similar instruments configuration), the CNN needs to be updated to account for different polarities in active regions or different heliocentric angles.

   \label{nnd}

\section{Python wrapper for the SIR, DeSIRe, and RH codes}
\label{warpper}
One of the easiest ways to boost inversions speeds is to simply take full advantage of modern computers, which usually have several or tens of CPUs, or use clusters. Going from a single-threaded program to a parallel version is often a laborious and intricate task, but in our case it can be done simply by inverting the independent pixels using separate threads. All the computations carried out in this paper were done using parallel computing. In this section we provide a general description of the parallel wrappers\footnote{Based on Python language. For more details and to download the source code and manual, go to: \url{https://gitlab.com/gafeira/parallel\_desire\_sir\_rh}} that we specifically developed for these tests. This will allow any potential user to start parallel inversions using the SIR code, DeSIRe (in preparation), or forward modelling using the non-LTE code RH \citep{2001ApJ...557..389U}. For instance, an SIR inversion of 370x370 pixels takes about one hour in a 60-core computer and almost half that time using the assisted inversion method.

The idea behind these scripts is to control the environment that each code needs for running an individual inversion in a pixel, distribute all pixels between different processors, and collect the results afterwards. This will allow the user to easily set up inversions, syntheses, and computations of response functions of large datasets, with single and multi-line computations from any instrument or line, in reasonable times. Moreover, it includes options in the SIR and DeSIRe wrappers for choosing different types of atmospheric model initializations: one single guess atmosphere for all the pixels; predefined clustering criteria and respective model atmospheres; starting from a previously inverted atmosphere; using several initial atmospheres and letting the code select the one that best fits the profiles; and CNN initialization.
In the last option, the user just needs to download the CNN model for the respective instrument and observed line (available in the wrapper git repository), and the code will automatically estimate the atmospheric initial condition for each pixel and run the inversion.

For example, the reference set took 4 hours and 8 minutes to run, and the assisted inversion took 1 hour and 45 minutes. Both inversions were performed on the same machine, which runs on 60 cores. 

The RH wrapper has the advantage that, compared to the other parallel  non-LTE radiative transfer codes (for example, RH 1.5D \cite{2015asclsoft02001P}), it allows the user to use the full capabilities and flexibility of the well-known RH code, keeping exactly the same structure of the main source code. The wrapper is also compatible with several versions of RH, permitting the users to maintain different versions of the main executable script.


\section{Summary}

In this paper we have presented a CNN implementation as a tool to provide initial guess atmospheric models to an inversion code, with a primary focus on parallel inversions. As a proof of concept, we applied this method to the Fe I 6301.5 and 6302.4 \AA{} pair of lines taken by the spectropolarimeter on board the Hinode spacecraft. We inverted the data using the SIR code plus the parallel wrapper described in Sect. \ref{warpper}.

In order to evaluate the performance of the CNN, we tested two different training approaches. The first approach uses just part of the map to train the CNN (red box in Fig. \ref{plot_full}). Finally, we used the full map to train the CNN in one go. For the two cases, we trained the CNN using 200 epochs. 

Using these two different training strategies, we identified some limitations of the CNN and its global performance. As shown in the scatter plots in Fig.~\ref{figbig}, the CNN performs well at estimating the physical parameters on the pixels that were used for training purposes but is less accurate in those areas of the map that were not used during the training process. 

This behaviour is a strong indication that the CNN model output needs to be analysed with extreme care.\ Therefore, the user should not blindly trust the results.

Based on these limitations, we used only the CNN estimate as input for further inversions. We tested this approach in the frame of parallel inversion with the goal of boosting the process and reducing computational times and costs. From the comparison of the CNN assisted inversion and the reference set (standard inversion) described in Sect. \ref{nnd}, we see some differences between them. This may be caused by the limitation of the method or by the uncertainties and degeneracies of the inversion method itself. We then ran a second reference inversion with a different node configuration but with a similar average $\chi^2$ and compared it with our first reference set. As can be seen in Table \ref{table_corr}, the dispersions of these two inversions are similar to the dispersion associated with the CNN assisted inversion. This indicates that the differences between the presented methods are likely not caused by the CNN assisted inversion, but are rather due to the inherent uncertainty in the inversion.

We also tested the assisted inversions in the scenario where a previously computed neural network model is used on a totally different dataset to estimate the initial atmosphere for the assisted inversion. As we see in Fig. \ref{qs_map} and in the scatter plot in Fig. \ref{qs_as}, the assisted inversion, even with a unseen dataset for the CNN model, shows a very high performance. In this particular case the performance of the assisted inversion, both in the time saved and the accuracy in retrieving the reference set values, is higher than in the case presented in Fig.~\ref{as_inv}. This may be caused by the simplicity of the profiles presented in this observation, which leads to a better constrain and consequently retrieve by the CNN of the atmospheric properties. That results in a higher performance of the assisted inversion because it is closer to the `real' atmosphere, making the inversion converge faster, saving time, and limiting the dispersion in the final results.

We can thus conclude that the assisted inversion can produce similar results to those of the reference inversions in a much shorter time, and can even improve the accuracy of the latter in some areas, such as cool umbrae. Therefore, we foresee CNN assisted inversions becoming a very helpful tool for initializing LTE, as well as non-LTE, inversions in the future.

\begin{acknowledgements}
We thank to Azaymi Siu Tapia on discussions and suggestions that greatly improved the manuscript.
All the network training and inference has been done using Keras with Tensorflow back-end. All the plots were done using matplotlib python package. This work has been supported by the Spanish Ministry of Economy and Competitiveness through projects ESP-2016-77548-C5-1-R and by Spanish Science Ministry ``Centro de Excelencia Severo Ochoa'' Program under grant SEV-2017-0709 and project RTI2018-096886-B-C51. D.O.S. also acknowledges financial support through the Ram\'on y Cajal fellowship. CQN acknowledges the Research Council of Norway through its Centres of Excellence scheme, project number 262622. CITEUC is funded by National Funds through FCT - Foundation for Science and Technology (project: UID/MULTI/00611/2019) and FEDER – European Regional Development Fund through COMPETE 2020 – Operational Programme Competitiveness and Internationalization (project: POCI-01-0145-FEDER-006922)
\end{acknowledgements}


\bibliographystyle{aa}
\bibliography{references.bib}

\begin{appendix}
\section{Assisted inversion result for the lines Fe I 15647.410 and 15648.515 \AA{}}

This section shows the results obtained from the assisted inversion following the strategy explained in Sect. \ref{nnd} and using data taken with the Gregor telescope in the 15647.410 and 15648.515 \AA{} lines. 
We used the full field-of-view to train the neural network, which took approximately 6 hours. 
Once again, the neural network was trained using a ground truth inversion that took approximately 4 hours, and the assisted inversion took around 1.9 hours. In this case it took less than a 1 minute for the initial atmospheres used for the assisted inversion to be estimated (including the interpolation).

\begin{figure*}
\centering
\includegraphics[width=\hsize]{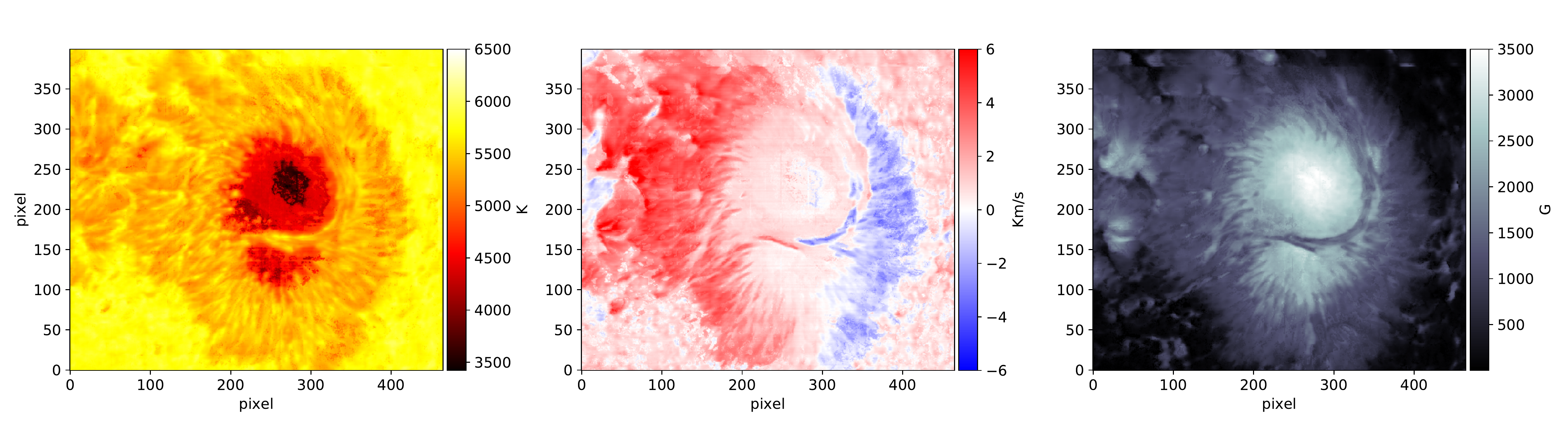}
\includegraphics[width=0.67\hsize]{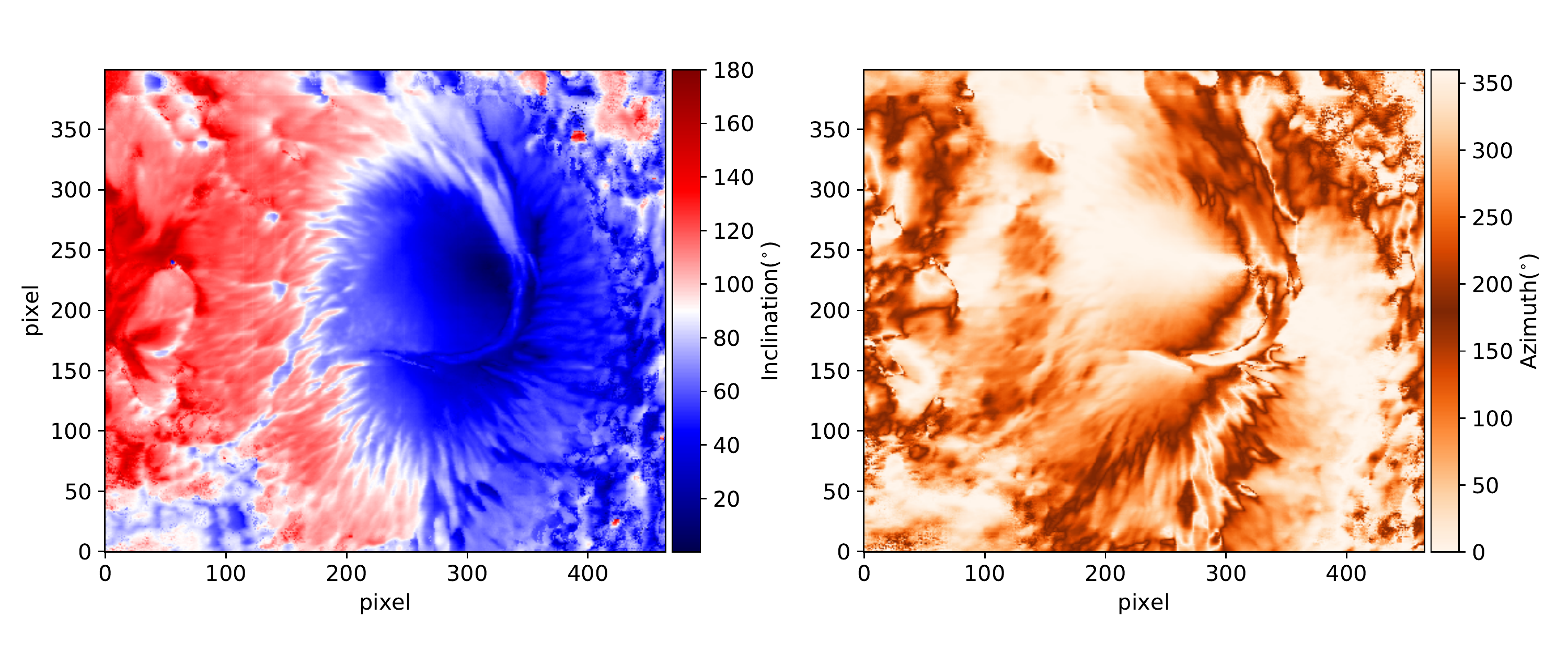}

\caption{Temperature, velocity, and magnetic field strength, inclination, and azimuth maps inferred from the SIR assisted inversion using Gregor data.}
 \label{gregor_ir_as}
\end{figure*}

\begin{figure*}[h]
\centering
\includegraphics[width=\textwidth]{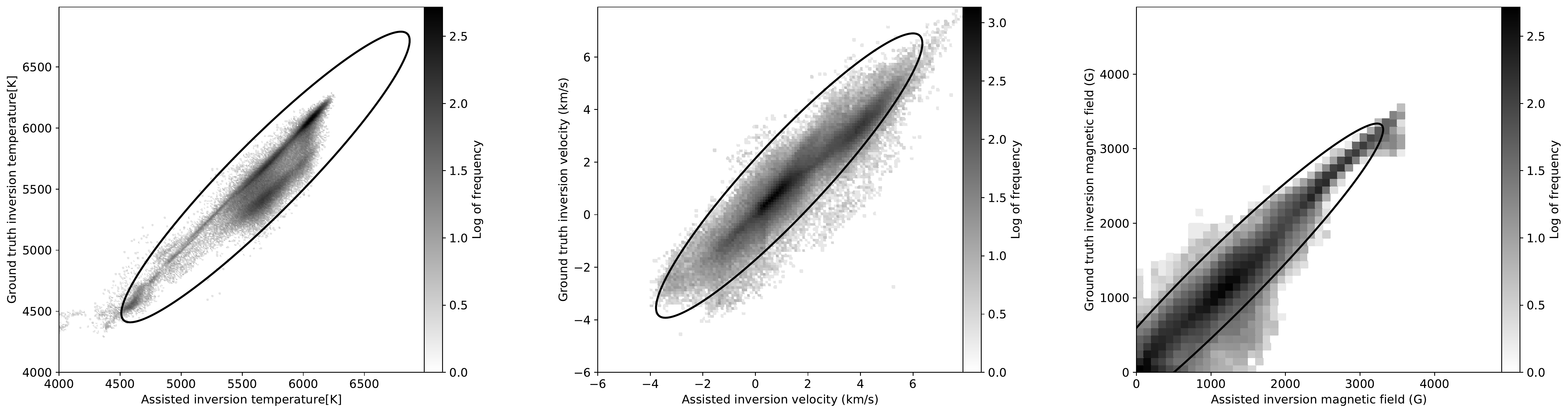}

\caption{Log scale of the 2D histogram of the temperatures, magnetic field strengths, and velocities obtained by the reference set inversion and the assisted inversion. The curved black line indicates the 99\% confidence level.}
 \label{ir_t_as}
\end{figure*}

In Figs. \ref{gregor_ir_as} and \ref{ir_t_as}, the resulting maps (similar to  Figs. \ref{as_inv} and \ref{t_as}) from the assisted inversion can be seen. In this case the assisted inversion performs as well as the previous example (with the Hinode data) shown in Sect. \ref{as_inv}.

\begin{table}[h]
\caption{Pearson correlation factor between the assisted and reference set inversions for each atmospheric physical parameter obtained from the Gregor data inversion.}
\label{table_corr_ir}
\begin{tabular}{cc}
 Atmospheric & Correlation\\ 
   physical parameter & Assisted inversion \\ \hline
 Temperature& 0.94\\
 Velocity& 0.93\\
 B strength &0.96\\
 B azimuth & 0.99\\
  B inclination&0.95\\ 
\end{tabular}
\end{table}
\label{ap2}
\end{appendix}

\end{document}